\documentstyle[prb,aps,multicol,epsf]{revtex}

\begin{document}

\newcommand{\etal}{{\em et al}.}
\newcommand{\EF}{$E_{\rm F}$}
\newcommand{\TK}{$T_{\rm K}$}
\newcommand{\rf}{$4d\rightarrow4f$}
\newcommand{\rt}{$3d\rightarrow4f$}

\widetext
\draft

\title{High-resolution Ce 3$d$-edge resonant photoemission study of CeNi$_2$}

\author{See-Hun Yang\cite{address1} and S.-J. Oh\cite{address2}}
\address{Department of Physics \& Center for Strongly Correlated Materials
Research,
Seoul National University, Seoul 151-742, Korea }

\author{Hyeong-Do Kim}
\address{Department of Physics, University of Seoul, Seoul 130-743, Korea}

\author{Ran-Ju Jung\cite{address3}}
\address{Japan Synchrotron Radiation Research Institute, Sayogun,
Hyogo 6779-5198, Japan\\
and Department of Material Physics, Osaka University, Osaka 560-8531, Japan}

\author{A. Sekiyama, T. Iwasaki, and S. Suga}
\address{Department of Material Physics, Osaka University, Osaka 560-8531, Japan}

\author{Y. Saitoh}
\address{Japan Atomic Energy Research Institute, SPring-8, 1-1-1, Koto,
Mikazuki, Sayo, Hyogo 079-5148, Japan}
\author{E.-J. Cho}
\address{Department of Physics, Chonnam National University, Kwangju 500-757,
Korea}

\author{J.-G. Park}
\address{Department of Physics, Inha University, Inchon 402-751, Korea}

\date{Received \hspace*{30mm}}

\maketitle

\begin{abstract}
Resonant photoemission (RPES) at the Ce \rt\ threshold has been performed for
$\alpha$-like compound CeNi$_2$ with extremely high energy
resolution
(full width at half maximum $<$ 0.2~eV) to obtain bulk-sensitive 4$f$
spectral weight.   The on-resonance spectrum
shows a sharp resolution-limited peak
near the Fermi energy which can be assigned to the tail of the Kondo
resonance.   However, the spin-orbit side
band around 0.3~eV binding energy corresponding to the $f_{7/2}$ peak is
washed
out, in contrast to the RPES spectrum at the Ce \rf\ threshold.      This
is interpreted as due to the different
surface sensitivity, and the bulk-sensitive Ce \rt\ RPES spectra are found
to be consistent with other electron
spectroscopy and low energy properties for $\alpha$-like Ce-transition
metal compounds, thus resolves
controversy on the interpretation of Ce compound photoemission. The 4$f$
spectral weight
over the whole valence band can also be fitted
fairly well with the Gunnarsson-Sch\"onhammer calculation of the single
impurity Anderson
model, although the detailed features show some dependence on the hybridization
band shape and (possibly) Ce 5$d$ emissions.
\end{abstract}


\begin{multicols}{2}
\narrowtext
%
%
For several decades Ce metal and its compounds have attracted much attention
because of their interesting physical properties such as Kondo behavior,
mixed valency, heavy fermion property, various magnetic states, and
superconductivity, etc.
Such properties are believed to originate from the interplay of strong
correlation
between Ce 4$f$ electrons and hybridization between 4$f$ and conduction
electrons, which is usually described by the periodic Anderson model.
Although it is now generally agreed  that low energy properties are well
described by
the Anderson model, there is still controversy as to the interpretation of
high energy probes such as photoemission
and inverse photoemission,\cite{allen,joyce} which directly measure
one-electron spectral weights.
Gunnarsson-Sch\"onhammer calculation (GS: Ref.~\onlinecite{gs}) and
noncrossing
approximation (NCA: Ref.~\onlinecite{nca}) of
an impurity version of the model, i.e., the single impurity Anderson model
(SIAM),
make it possible to compare directly  theoretical 4$f$-electron spectrum
with experimental photoemission data.
Thus in principle one can obtain model parameters of the SIAM for each
compound from photoemission data,
which can then be used to understand its low-energy properties.
Resonant photoemission spectroscopy (RPES) at the Ce \rf\ edge,
x-ray photoelectron spectroscopy (XPS) for Ce 3$d$ core-levels,
and bremsstrahlung isochromat spectroscopy (BIS) have been used for
this purpose and shown to be quite successful
for many Ce compounds.\cite{allen}
On the other hand, Arko and co-workers\cite{joyce} dispute this interpretation,
claiming that the 4$f$ weights
of many Ce compounds measured by photoemission do not follow
these schemes in
that \rf\ RPES spectra of extremely $\alpha$-like Ce compounds show some
discrepancy
with core-level XPS and BIS spectra, which has not been completely understood
as yet.

One possible source of these discrepancy and controversy is the surface
effect.
From angle-dependent Ce 3$d$ core-level XPS spectra and
threshold-dependent RPES spectra of several $\alpha$-like Ce compounds,
Laubschat \etal\ proposed that surface electronic structures of
those compounds are not $\alpha$-like but $\gamma$-like,\cite{laubschat}
which is now pretty well established.\cite{duo}
Since the photon energy of the \rf\ threshold is so low that
\rf\ RPES is quite surface sensitive, the discrepancy between experimental
\rf\ RPES spectrum and theoretical one, which is obtained from parameters
mainly determined by XPS and BIS,
can be understood in terms of surface effects.
In this context, \rt\ RPES is more desirable to examine bulk
electronic structures of Ce compounds because the escape depth of
photoelectrons is longer.
However, the resolution of  photon source around the \rt\ threshold has
been much poorer than that at the
\rf\ threshold, which rendered limited information.

In this work, we present \rt\ RPES spectra of very high Kondo temperature 
material CeNi$_2$ ($T_{\rm K} \sim 1000$~K) with the extremely high
experimental energy resolution (0.2~eV full width at half maximum
(FWHM)).\cite{saitoh}
We found that on-resonance spectrum shows a sharp
resolution-limited peak near the Fermi
energy (\EF) which can be assigned to the tail of the Kondo resonance.
Comparison with a GS calculation of the SIAM shows good agreement between
theory and experiment, thus high-resolution \rt\ RPES opens new
opportunities
to study {\em bulk} electronic structures of Ce compounds.

%
%

Polycrystalline CeNi$_2$ was prepared by arc melting of high-purity metals
under argon atmosphere.
The structure and homogeneity were checked by x-ray diffraction.
\rt\ RPES measurements of CeNi$_2$ were performed at the beamline BL25SU of
the SPring-8.   FWHM of photon source around the \rt\
threshold was better than 200~meV and the temperature of the sample was
maintained at 30~K throughout the measurements.
The SCIENTA SES200 electron analyzer was used to obtain an overall
experimental
resolution of $\sim$~0.2~eV FWHM.
Clean sample surface was obtained by scraping {\it in situ}
with a diamond file under the pressure of $4 \times 10^{-10}$~Torr.
\EF\ of the sample was referenced to that of a gold film
deposited onto the sample substrate.
\rf\ RPES measurements of CeNi$_2$ were also carried out at the beamline
BL-3B of
the Photon Factory, High Energy Accelerator Research Organization
(KEK) in Tsukuba. FWHM of photon sources around the \rf\
threshold was about 30~meV and the overall experimental resolution of
40~meV 
was obtained with the SCIENTA SES200 electron analyzer.
Scraping was incorporated for the sample cleaning under the base pressure
better than $5\times 10^{-10}$~Torr, and all the measurements were done at
30~K.
\EF\ of the sample was referenced to that of a gold film
deposited onto the sample substrate and its position was accurate to better
than 2 meV.

%
%

Figure 1 shows the valence-band RPES spectra of CeNi$_2$ around the Ce
\rf\ threshold.
All the spectra are normalized according to the photon flux.
The spectra are overall consistent with early data \cite{allen}\ except
for the difference of energy resolution.
As the photon energy changes, the spectrum does not show a remarkable
resonant enhancement of the Ce 4$f$ character in contrast to other
Ce-non-transition-metal compounds.
This fact was already noticed in the previous poorer-resolution \rf\ RPES
study and was attributed to strong Ni 3$d$ emission.\cite{allen}
Photoemission cross section of Ni 3$d$ electrons
is strongly dependent on the photon energy around the \rf\
threshold,\cite{yeh}
thus it is hardly possible to extract reliable Ce 4$f$ removal spectrum
using the conventional method.\cite{comment1}

\begin{figure}
\epsfxsize=65mm
\centerline{\epsffile{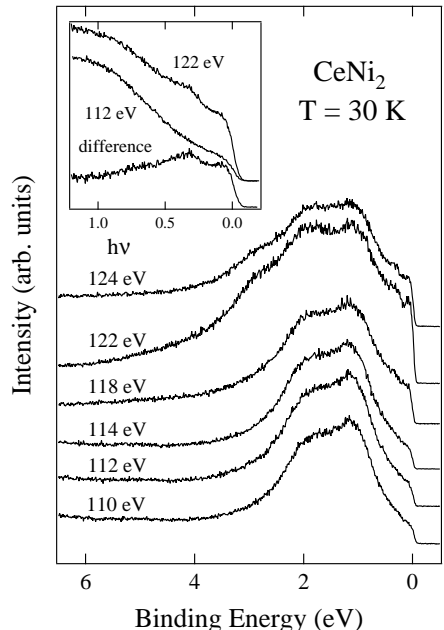}}
\caption{Valence-band \rf\ RPES spectra of CeNi$_2$ at $T = 30$~K.    Inset
shows detailed spectra near \EF.}
\end{figure}

In the on-resonance spectrum at $h\nu = 122$~eV, we can see
that two features grow up at about 3~eV and near \EF.
The former could be assigned to an $f^0$ peak, and the latter to an $f^1$ one.
In the inset of Fig.~1,  the detailed spectra of  the $f^1$ peak in the narrow
region near \EF\ is shown.
Similar to other Ce compounds, two features are enhanced on resonance.
As usual, we can assign
the peak at the Fermi level to the tail of Kondo resonance of the
$f_{5/2}$ peak, while the one
around 0.3~eV binding energy is its spin-orbit side band from the
$f_{7/2}$ peak.   The fact that $f_{7/2}$ side band
is clearly observed around 0.3~eV binding energy is somewhat inconsistent
with the GS analysis (see below).


\rt\ RPES spectra of CeNi$_2$ are presented in Fig.~2.
Contrary to the case of \rf\ RPES in Fig.~1, the Ce 4$f$ character is
dramatically enhanced
in the on-resonance spectrum ($h\nu=881.4$~eV) in comparison with the
off-resonance spectrum ($h\nu=868.1$~eV).\cite{comment2}
Especially, thanks to the extremely high resolution,
we can see a very sharp peak at \EF, whose position is limited by
the experimental resolution.   Thus this peak is undoubtedly assigned to the
tail of the Kondo resonance as was done for a lower-\TK\ CeSi$_2$ system.\cite{garnier}
We also observe a small hump around 1~eV binding energy and a broad
feature around 3~eV binding energy.
The broad feature around 3~eV binding energy probably originates from the $f^0$
character as generally accepted, but the origin of the 1~eV peak is a little
controversial\cite{lawrence,kim} and this will be discussed later.

\begin{figure}
\epsfxsize=65mm
\centerline{\epsffile{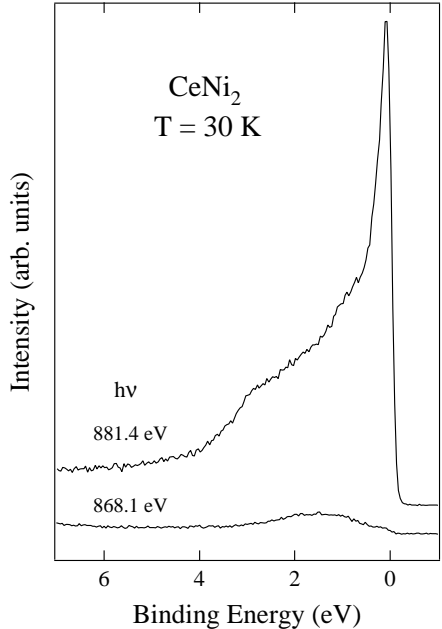}}
\caption{Valence-band \rt\ RPES spectra of CeNi$_2$ at $T = 30$~K.}
\end{figure}

Another interesting point is that we do not see any structure around
0.3~eV binding energy in the \rt\ on-resonance spectrum, which corresponds to
the $f_{7/2}$ peak and is clearly noticeable in \rf\ RPES of Fig~1.
We first suspected this may be due to the poorer energy resolution of \rt\ RPES than
that of \rf\ RPES,
but we discarded this possibility for the following reason.
In order to see whether the lineshape is due to the experimental
resolution, we simulated a 4$f$ spectrum of a low-\TK\ system, in which
the $f_{7/2}$ peak is clearly resolved with 0.1~eV resolution
(FWHM),\cite{comment3} with our experimental resolution determined
by fitting gold \EF\ spectrum.
We then found that the highest peak position is around
the center of the $f_{5/2}$ and $f_{7/2}$ peaks and the lineshape is
rather symmetrical.
These facts contradict the on-resonance spectrum
in that the highest peak position
is very close to \EF\ and the lineshape is quite asymmetric
as shown in Fig.~2,
which implies that the intensity of the $f_{7/2}$ peak is smaller than
in the \rf\ RPES spectrum or the peak is indistinguishable from the
tail of the Kondo resonance.
In fact, according to the GS and NCA schemes of the SIAM,
the lineshape of the $f_{7/2}$ peak shows such a behavior
as \TK\ increases.\cite{malterre}
We conclude that the spin-orbit side band observed in previous
high-resolution \rf\ RPES and
He {\footnotesize II} photoemission spectra of high-\TK\ Ce compounds,
which was not well reproduced by GS and NCA calculations with parameters
suitable for bulk physical properties,
originates from the surface
where the Ce 4$f$ spectrum is more $\gamma$-like.
This fact was also noticed by Kim \etal\ by analyzing \rf\ and \rt\ RPES
spectra of CeIr$_2$.\cite{kim}

\begin{figure}
\epsfxsize=65mm
\centerline{\epsffile{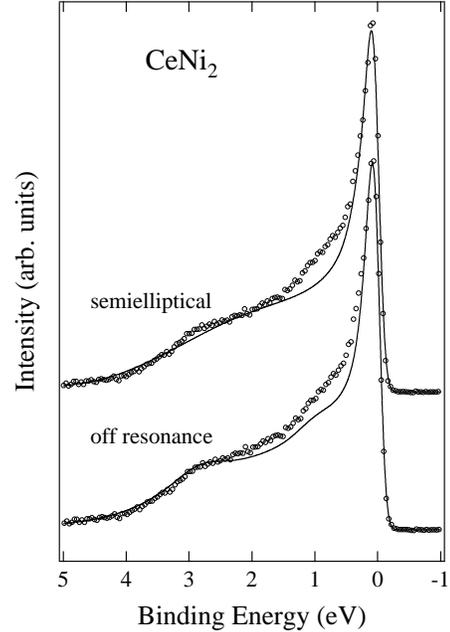}}
\caption{
Comparison of GS calculations (solid lines) of the SIAM with
the Ce 4$f$ spectrum (empty circles) of CeNi$_2$ derived from \rt\ RPES spectra.
The lower graph shows GS calculation using the off-resonance spectrum
for $V(\varepsilon)^2$, the upper graph a semielliptical shape.
For parameter values and detailed procedure, refer to text.}
\end{figure}

In order to see whether the bulk-sensitive 4$f$ spectrum obtained from 
\rt\ RPES of CeNi$_2$ is quantitatively explained by the SIAM,
we have performed GS calculations which includes spin-orbit
splitting of the 4$f$ level.
Since it is not simple to separate surface and bulk contribution from
the experimental data, we will neglect the surface effect 
for the bulk-sensitive \rt\ RPES spectra here.
Figure~3 shows the 4$f$ spectrum derived from \rt\ RPES spectra (empty circles)
and the GS-calculation results (solid lines) employing the \rf\ off-resonance
spectrum (lower graph) and a semielliptical shape (upper graph) 
for the hybridization matrix elements $V(\varepsilon)^2$.
For basis states we employed the lowest order $f^0$, $f^1$, and $f^2$, and
the second-order $f^0$ states.
The used parameter values are as follows: The 4$f$-electron energy
$\varepsilon_f$ is $-1.13$,
the spin-orbit splitting of the $f$ level $\Delta_{so}$ is 0.28~eV,
the hybridization strength averaged over the occupied valence band
$\Delta_{av}$ is 89~meV,
and the on-site Coulomb interaction between 4$f$ electrons $U$ is 6~eV,
which give the 4$f$-level occupancy $n_f = 0.78$.
The static, $T=0$ susceptibility $\chi(0)$ of CeNi$_2$ 
gives the estimates $n_f = 0.76$ and 0.83 depending on the reference
compound,\cite{sakurai} which is comparable to the present
spectroscopic estimate.
To compare the theoretical spectrum with experimental data, we first
broadened the calculated spectrum
with a Lorentzian of the width given by $0.01+0.20|E-E_{\rm F}|$~eV, and then
the spectral weight above \EF\ was removed using the method of
Liu \etal,\cite{liu} and finally the resulting curve was convoluted by a
Gaussian for experimental resolution.

The theoretical curves shown in Fig.~3 match the
experimental data quite well, especially near the \EF\ region and the
bottom of the valence band.
This is taken as the evidence that the GS calculation with parameter
values consistent with low
energy properties can reproduce the experimental photoemission spectra
well even for high-\TK\ material CeNi$_2$.
The only region showing discrepancy between theory and experiment is
around the binding energy of 1~eV.
Similar 1~eV structure has been observed before in other Ce compounds, and
its origin was a little
controversial.\cite{lawrence,kim}
Lawrence \etal\cite{lawrence} claimed that the contribution of Ce 5$d$ emission,
whose position is around 1~eV, to the 4$f$ spectrum is considerable (about
30~\%).
Recent angle-resolved RPES studies of LaSb (Ref.~\onlinecite{olson})
and La metal (Ref.~\onlinecite{molodtsov}) show enhancement of La 5$d$ emission
at the \rf\ resonance, 
although its magnitude is much less than claimed in Ref.~\onlinecite{lawrence}.
The enhancement of La 5$d$ emission in La compounds was also
observed in \rt\ RPES.\cite{suga}
On the other hand, such 1~eV structure could be reproduced by the GS calculations
without considering 5$d$ emission as demonstrated in the cases of
$\alpha$- and $\gamma$-Ce metal using
realistic hybridization shape $V(\varepsilon)^2$ by Liu \etal,\cite{liu}
and recently it was also proposed that similar 1~eV structure for CeIr$_2$
would be reproduced if realistic $V(\varepsilon)^2$ is used in the
GS calculations.\cite{kim}
Though the off-resonance spectrum may not be very realistic for
$V(\varepsilon)^2$,
our GS calculation presented in Fig.~3 using this
off-resonance curve reveals distinctive 1~eV structure,
which is not observed in the calculation using structureless
semielliptical band, where all other parameter values
are kept the same (upper graph of Fig.~3).
It strongly suggests that the hybridization between Ce 4$f$ and Ni 3$d$
electrons plays an important role for this 1~eV peak, although 5$d$
emission may also contribute. 
Thus it is quite essential to employ realistic
$V(\varepsilon)^2$ in GS calculations in order to fully interpret
experimental spectra.

%
%
In conclusion, we have performed high-resolution \rf\ and \rt\
RPES measurements of CeNi$_2$.
It was nearly impossible to extract Ce 4$f$ spectrum from the \rf\
RPES spectra because of overlapping Ni 3$d$ bands, but the \rt\ RPES spectra
with extremely high resolution
provide a clear bulk-sensitive 4$f$ spectrum.
The experimental 4$f$ spectrum thus obtained is well reproduced using a GS
calculation
of the SIAM.

\vspace*{4mm}
This work is supported by the Korean Science and Engineering Foundation
(KOSEF) through the Center for Strongly
Correlated Materials Research (CSCMR) at Seoul National University (1999),
and Grant-in-Aid for COE Research (10CE2004) of the Ministry of Education,
Science, Sports and Culture, Japan.
The authors thank K. Matsuda, M. Ueda, H. Harada, T. Matsushita, M. Kotsugi,
and T. Nakatani for partial support for the experiments.
The 3$d$ RPES was performed under the approval of the Japan Synchrotron
Radiation Research Institute (1997B1047-NS-np).
The 4$d$ RPES was performed under the approval of PF-PAC (92S002).

\end{multicols}

\begin{references}
\bibitem[\dagger] {address2}
Author to whom all correspondences should be addressed.
\bibitem{comment1}
We tried to extract surface and bulk Ce 4$f$ spectra from the \rf\ and
\rt\ RPES spectra using the conventional method. 
The derived surface spectrum, however, shows
strong negative intensity near \EF\, probably due to the inappropriate
extraction
of Ce 4$f$ spectrum at the \rf\ threshold.
\bibitem{comment2}
The off-resonance spectrum was taken at somewhat lower photon energy, because
around the usual value (875~eV: Ref.~\onlinecite{duo}) there is strong
enhancement
of the Ni 3$d$ character due to Ni 2$p_{1/2}\rightarrow 3d$ resonant process.
\bibitem{comment3}
We could indeed observe the well-resolved $f_{7/2}$ peak in the \rt\
RPES of low-\TK\ systems, when we improved the resolution to 0.1~eV.
\end{references}
\end{document}